\documentclass[9pt,twocolumn,twoside]{osajnl}
\usepackage{upgreek}

\journal{ol} 

\setboolean{shortarticle}{true}

\title{Cooling neutral atoms into maximal entanglement in the Rydberg blockade regime}

\author[1]{Wei-Lin Mu}
\author[1,3]{Xiao-Xuan Li}
\author[1,2,4]{Xiao-Qiang Shao}

\affil[1]{Center for Quantum Sciences and School of Physics, Northeast Normal University, Changchun, 130024, China}
\affil[2]{Center for Advanced Optoelectronic Functional Materials Research,
and Key Laboratory for UV Light-Emitting Materials and Technology of Ministry of Education,
Northeast Normal University, Changchun 130024, China}
\affil[3]{lixx083@nenu.edu.cn}
\affil[4]{shaoxq644@nenu.edu.cn}

\begin{abstract}
We propose a cooling scheme  to prepare stationary entanglement of neutral atoms in the
Rydberg blockade regime by combination of periodically collective laser pumping and dissipation. In each cycle, the controlled unitary dynamics process can selectively pump atoms away from the non-target state while maintaining the target state unchanged. The subsequent dissipative process redistributes the populations of ground states through the engineered spontaneous emission. After a number of cycles, the system will be eventually stabilized into the desired steady state 
independent of the initial state. This protocol does not rely on coherent addressing of individual neutral atoms or fine control of Rydberg interaction intensity, which can in principle  greatly improve the feasibility of experiments in related fields.
\end{abstract}

\setboolean{displaycopyright}{true}

\begin{document}
\maketitle
Quantum decoherence, resulting from the inevitable interaction between the interested quantum system and its surroundings, has recently been reexamined as a resource in quantum information processing. 
The earliest idea of using dissipation in generation of entangled states dates back to the pioneering work made by Plenio \emph{et al.}, where they provided a probabilistic scheme for generating
an entangled state of two atoms by continuous monitoring of photons
leaking out of the cavity \cite{1}. After that, the nontrivial dissipative interactions combined with continuous and stationary control fields are exploited to stabilize entangled states on various physical platforms in a deterministic manner \cite{2,3,4,5,6,7,8,9,10,11}. Neutral-atom systems are considered to be powerful candidates among many physical systems for quantum information processing because the highly excited Rydberg states possess long lifetimes and extremely large dipole moments, especially the latter feature can suppress the double excitation of Rydberg states from ground-state atoms over a long distance relative to  the atomic scale \cite{12}. In contrast with Rydberg blockade, the Rydberg antiblockade has more advantages in dissipative preparation of entangled states due to the selective excitation of ground states \cite{13,14,15}. Nevertheless, such kind of facilitation dynamics requires a delicate balance between the Rydberg interaction intensity and the single-photon detuning parameter.
Although the unconventional Rydberg pumping mechanism proposed by our group can reduce the influence of interatomic distance fluctuation to a certain extent, it is still not applicable for preparation of multi-atom entanglement by dissipation \cite{16,17}. In this letter, we intend to incorporate the Rydberg blockade mechanism into the  dissipative protocols, such as multi-dimensional entanglement and multipartite entanglement, which brings an advantage that the scheme becomes first-order independent of the large blockade shift, so the defects in the Rydberg antiblockade-based schemes can be avoided. Meanwhile, in order to solve the problem of the stationarity of the target state under driving, we introduce the pulsed laser instead of continuous laser, and only  require the target state to remain stable  at the end of the pulse node, rather than the entire driving process. The aforementioned unitary dynamics combined with the atomic spontaneous emission constitutes a cycle of our scheme, and after 
many cycles, the system will be stabilized into the maximally entangled state irrespective to the initial state of system. This method can be described mathematically as floquet-Lindblad dynamics and physically as modified laser sideband cooling. 
It is noteworthy that the idea of pumping cycles has only ever been used to improve continuous implementation in trapped-ion systems \cite{18,19,20,21}, while the alternate driving of various pulses is a requirement for the current model due to the fact that continuous laser driving is unable to provide selective excitation for the ground states in the Rydberg blockade regime.

\begin{figure}
    \centering
    \includegraphics[scale=0.1862]{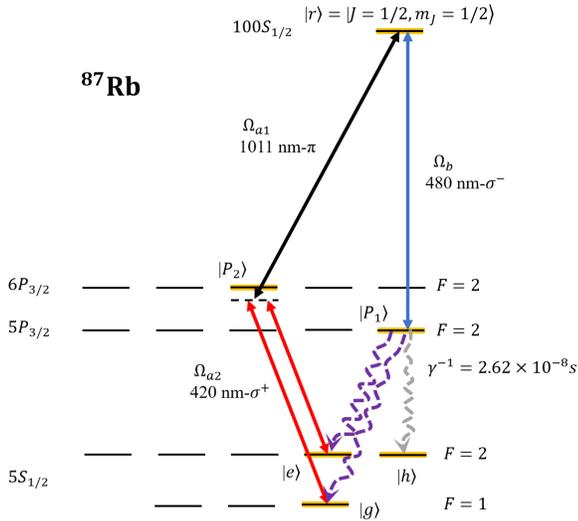}
    \caption{The internal states of atoms. This model
    incorporates three ground states $|g\rangle$, $|e\rangle$, and $|h\rangle$ of the  Zeeman sublevels of the $5S_{1/2}$,  temporary (short-lived) levels $|p_1\rangle$ of $5P_{3/2}$ and $|p_2\rangle$ of $6P_{3/2}$, and a Rydberg state $|r\rangle$ of $100S_{1/2}$. Laser pulses of Rabi frequency $\Omega_{a}$ interacts resonantly with the
transition $|g\rangle\leftrightarrow|r\rangle$ ($|e\rangle\leftrightarrow|r\rangle$) and 
laser pulses of Rabi frequency $\Omega_b$ drive the transition $|r\rangle\leftrightarrow|p_1\rangle$.  The lifetime of the short-lived state $|p_1\rangle$ is $\gamma^{-1}=2.62\times10^{-8}s$.}
    \label{fig1}
\end{figure}

The relevant level structure of $^{87}$Rb atom is shown in Fig.~\ref{fig1}. Three ground states $|g\rangle=|F=1,m_F=1\rangle$, $|e\rangle=|F=2,m_F=1\rangle$, and $|h\rangle=|F=2,m_F=2\rangle$ are the Zeeman sublevels of $5S_{1/2}$. The intermediate states $|p_1\rangle=|F=2,m_F=2\rangle$ and $|p_2\rangle=|F=2,m_F=0\rangle$ are temporary (short-lived) levels of $5P_{3/2}$ and $6P_{3/2}$, respectively. $|r\rangle=|J=1/2,m_J=1/2\rangle$ is the  Rydberg level of $100S_{1/2}$. $|g\rangle$ ($|e\rangle$) is driven to $|p_2\rangle$ by a $\sigma^+$ polarized 420~nm laser of Rabi frequency $\Omega_{a2}$, which is then pumped to the Rydberg state by a $\pi$ polarized 1011~nm laser of Rabi frequency $\Omega_{a1}$. The intermediate state $|p_2\rangle$ can be eliminated adiabatically under large detuning conditions, resulting in an effective pumping process driving the transition $|g\rangle\leftrightarrow|r\rangle$ ($|e\rangle\leftrightarrow|r\rangle$) with Rabi frequency $\Omega_a=\Omega_{a1}\Omega_{a2}/(2\Delta)$ \cite{22,23}, where $\Delta$ is the detuning parameter of the pumped lasers. 
In addition, we introduce another $\sigma^-$ polarized 480~nm laser pulse of Rabi frequency $\Omega_b$ to resonantly couple the Rydberg state $|r\rangle$ to the fast decaying state $|p_1\rangle$ of lifetime $\gamma^{-1}=2.62\times10^{-8}s$, so as to modulate the spontaneous emission rate of the Rydberg state. According to the selection rule for electric dipole transition, the branching ratios of decay from $|p_1\rangle$ to $|g\rangle$, $|e\rangle$, and $|h\rangle$ are 1/6, 1/2, and 1/3, respectively.
\begin{figure*}
\centering
    \includegraphics[scale=0.2]{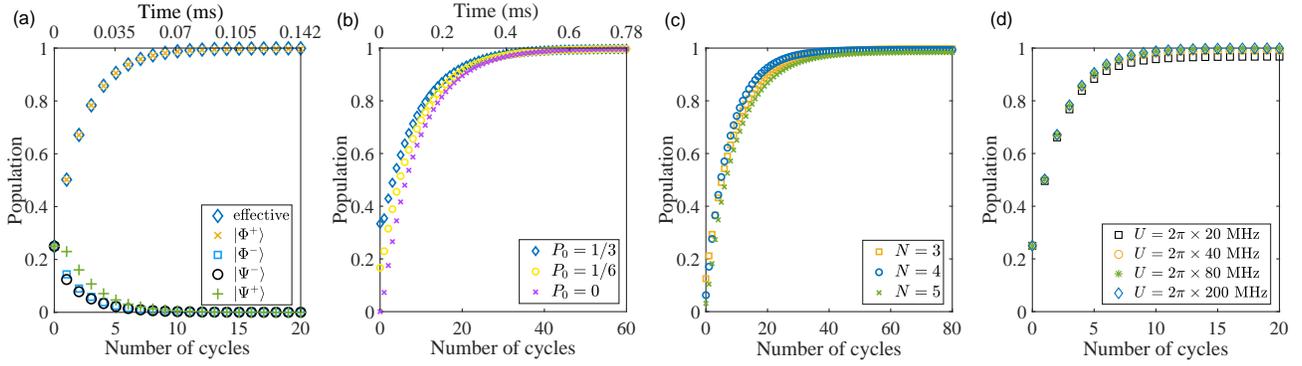}
    \caption{(a) Population evolution of different ground states with the number of cycles, where $|\Phi^-\rangle=1/\sqrt{2}(|gg\rangle-|ee\rangle)$, $|\Psi^+\rangle=1/\sqrt{2}(|eg\rangle+|ge\rangle)$, and $|\Psi^-\rangle=1/\sqrt{2}(|eg\rangle-|ge\rangle)$. 
    The population of the target state $|\Phi^+\rangle$ under the effective master equation is shown by the rhombus symbol, and the full dynamical evolution, including the natural spontaneous emission of the Rydberg state, is represented by the cross symbol.
     (b) Process of system evolution to the three-dimensional entangled state from three different initial mixed states. (c) Population evolution of different GHZ states with the number of cycles. (d) Population evolution of 
   $|\Phi^+\rangle$ under different Rydberg interaction intensities $U$. The relevant parameters are $\Omega_a=2\pi\times2$ MHz, $\Omega_b=2\pi\times1.2$ MHz, $\Omega_c=2\pi\times20$ kHz, and $\delta =2\pi\times\sqrt{{2}/{3}}$ MHz. }
    \label{fig2}
\end{figure*}

In order to better understand the principle of the scheme, we first discuss the dissipative generation of the Bell state of two atoms with form of $|\Phi^+\rangle=1/\sqrt{2}(|gg\rangle+|ee\rangle)=1/\sqrt{2}(|++\rangle+|--\rangle)$, where $|\pm\rangle={1}/{\sqrt{2}}(|g\rangle\pm|e\rangle)$. The whole process consists of repeated cycles with each consisting of three steps. In step A, we switch on the global laser driving the transition $|g\rangle\leftrightarrow|r\rangle$. Under condition where the Rydberg interaction $U$ is much stronger than the effective Rabi frequency $\Omega_a$, the effective Hamiltonian in the interaction picture can be described as 
$
\hat{H}_{a}^I={\Omega_a}/{2}(|er\rangle\langle eg|+|re\rangle\langle ge|)+\Omega_a/{2}(|rg\rangle+
|gr\rangle)\langle gg|+{\rm H.c.}.
$
  For atoms initialized in the ground states, the population of $|gg\rangle$ is recovered after a driving period $\tau_a={\sqrt{2}\pi}/{\Omega_a}$, while the populations of $|eg\rangle$ and $|ge\rangle$ cannot be completely restored due to the difference in the Rabi frequency between the single and collective excitation processes. Then we switch off the pump laser and start the dissipation process by 
 switching on the laser that drives the transition $|r\rangle\leftrightarrow|p_1\rangle$. 
 Under the premise of ignoring the natural Rydberg spontaneous emission, the engineered
spontaneous emission rate of $|r\rangle$ is derived as $\gamma_{\rm eff}=\Omega_b^2/\gamma$ \cite{17} when $\Omega_b\ll\gamma$. The corresponding Markovian master equation reads
$
\dot{\hat{\rho}} =\sum_{j=1}^2\lbrace{\gamma^g_{\rm{eff}}}/{2}\mathcal{\hat{L}}[|g\rangle_{jj}\langle r|]+{\gamma^e_{\rm{eff}}}/{2}\mathcal{\hat{L}}[|e\rangle_{jj}\langle r|]\rbrace\hat{\rho},
$ where $\mathcal{\hat{L}}[\hat{c}]\hat{\rho}=2\hat{c}\hat{\rho}\hat{c}^+-\lbrace\hat{c}^+\hat{c}\hat{\rho},\hat{\rho}\hat{c}^+\hat{c}\rbrace$. 
This dissipation process can redistribute the population remaining on $|r\rangle$ after executing the previous unitary dynamics to
$|g\rangle$ and $|e\rangle$ irreversibly. Note that the population of hyperfine ground state $|h\rangle$ is ignored here because it can be recycled with a series of $\pi$-polarized lights.
The operations of step B and step C are similar to that of step A, which both require controlled unitary evolution and dissipation. The only difference is that in step B, we apply the drive from $|r\rangle$ to $|e\rangle$, and in step C, we apply the drive from $|r\rangle$ to $|+\rangle$. The corresponding times should ensure that the population of $|ee\rangle$ and the probability amplitude of $|++\rangle$ each undergo a periodic oscillation, respectively.
Therefore, the system will be stabilized into $|\Phi^+\rangle$ after continuously cyclic evolution according to steps A, B and C.
Fig.~\ref{fig2}(a) shows that after 20 cycles ($0.142$ms), the population of the Bell state $|\Phi^+\rangle$ will be greater than 99$\%$ starting from the fully mixed state $\rho_0 =\sum_{i,j=g,e}|ij\rangle\langle ij|/4$, where the parameters are $U=2\pi\times400$ MHz, $\Delta=2\pi\times1250$ MHz, $\Omega_{a1}=2\pi\times50$ MHz, $\Omega_{a2}=2\pi\times50$ MHz, $\Omega_b=2\pi\times1.2$ MHz, $\gamma=2\pi\times6$ MHz, and the relaxation time in each step is chosen as $2\mu$s.

Based on the above scheme of preparing Bell state, the stationary three-dimensional entanglement can be produced by introducing microwave fields $
\hat{H}_{mw}^I={\Omega_c}/{2}\sum_{j=1}^{2}(|e\rangle_{jj}\langle g|+|h\rangle_{jj}\langle g|)+{\rm H.c.}
$ to globally drive hyperfine level transition \cite{PhysRevA.89.052313}.
Each cycle of this scenario only consists of step A' and step B',  which are similar to step A and step B in the preparation of Bell state, but the driving time should ensure that the probability amplitudes of $|gg\rangle$ ($|ee\rangle$) experiences a periodic evolution rather than the population, and the above microwave fields need to be applied after the engineered dissipation process characterized by
$
\dot{\hat{\rho}}=\sum_{j=1}^2\lbrace\gamma^g_{\rm{eff}}/2\mathcal{\hat{L}}[|g\rangle_{jj}\langle r|]+{\gamma^e_{\rm{eff}}}/{2}\mathcal{\hat{L}}[|e\rangle_{jj}\langle r|]+{\gamma^h_{\rm{eff}}}/{2}\mathcal{\hat{L}}[|h\rangle_{jj}\langle r|]\rbrace\hat{\rho}.
$ 
There are three degenerate dark states for $\hat{H}_{c}^I$, i.e., $|T_1\rangle={1}/{\sqrt{3}}(|ee\rangle-|gg\rangle+|hh\rangle)$, $|T_2\rangle={1}/{\sqrt{3}}(|eh\rangle-|gg\rangle+|he\rangle)$, and $|T_3\rangle={1}/{2}(|eg\rangle-|ge\rangle-|gh\rangle+|hg\rangle)$, of which the pump lasers of step A' and step B' can destroy all states except $|gg\rangle,|ee\rangle$, and $|hh\rangle$,  so the system will be stable in $|T_1\rangle$ after continuously cyclic evolution according to steps A' and B'. Fig.~\ref{fig2}(b) depicts the process of system evolution to the target state with three different initial mixed states $1/3(|gg\rangle\langle gg|+|ee\rangle\langle ee|+|hh\rangle\langle hh|)$, $1/6(|gg\rangle\langle gg|+|ee\rangle\langle ee|+|hh\rangle\langle hh|+|eg\rangle\langle eg|+|ge\rangle\langle ge|+|eh\rangle\langle he|)$, and $1/2(|eg\rangle\langle eg|+|ge\rangle\langle ge|)$. For all cases, after 60 cycles ($0.78$~ms), the population of the target state can reach more than 99\%, where we have chosen $\Omega_c=2\pi\times20$~kHz, $\tau_c={2\pi}/{(7\Omega_c)}$, and other parameters are the same as in the previous preparation of Bell state.
 
A prominent advantage of our current model over the  antiblockade effect and unconventional Rydberg pumping is that it can be directly extended to the preparation of multi-atom GHZ states. We decompose each cycle of preparing an $n$-atom GHZ state into step A'', step B'', and step C''.
The steps A'' and B'' are similar to steps A and B. The coupling strength between all atoms in $|g\rangle$ $(|e\rangle)$ and the corresponding  single excited collective
state is $\sqrt{n}\Omega_a/2$, thus the selection of time $2\pi/(\sqrt{n}\Omega_a)$ ensures that  when $n$ atoms are initially in the ground state $|g\rangle(|e\rangle)$, this probability will recover at the end of the pulse. 
 The duration of laser pulses applied in step C'' must be chosen in such a way that the probability amplitudes of states with even number of $|+\rangle$ undergo a periodic evolution.
  This process can be easily realized in the preparation of triatom GHZ state ${1}/{\sqrt{2}}(|ggg\rangle-|eee\rangle)={1}/2(|---\rangle+|++-\rangle+|+-+\rangle+|-++\rangle)$,
 because
 the duration of laser pulses driving $|+\rangle$ to $|r\rangle$ is the same as step C. 
 But for preparing GHZ states of four and five atoms such as $1/(2\sqrt{2})\sum_{k=0}^2|\mathcal{{X}}_{2K}^4\rangle$ and $1/4\sum_{k=0}^2|\mathcal{{X}}_{2K}^5\rangle$ with $|\mathcal{{X}}_{m}^n\rangle$ being the state incorporating $m$ atoms in $|+\rangle$ and $(n-m)$ atoms in  $|-\rangle$, the resonant lasers used in step C'' of preparing three-atom GHZ state cannot make the probability of amplitudes of states $|\mathcal{{X}}_{2}^n\rangle$ and $|\mathcal{{X}}_{4}^n\rangle$ restored simultaneously, since the ratio of the Rabi frequencies coupled the above states to the single excited Rydberg state is $1:\sqrt{2}$ rather than an integer ratio. Fortunately, we can achieve our goal by introducing two-photon-detuned pump lasers driving the transition $|r\rangle\leftrightarrow|+\rangle$. 
 The corresponding Hamiltonian reads
$\hat{H}_{c''}^I={{\sqrt{2}}/{2}\Omega_ae^{-i\delta t}}\sum_{j=1}^{n}(|r\rangle_{jj}\langle +|+|+\rangle_{jj}\langle r|)+\sum_{i<j}U_{i,j}|r_ir_j\rangle\langle r_ir_j|.$
 If the system is in the state of $|\mathcal{{X}}_{m}^n\rangle$ initially, the probability amplitude that there still $m$ atoms in $|+\rangle$ at time $t$ is ${(P_m^+e^{P_m^+t}-P_m^-e^{P_m^-t})}/{iW_m}$, where $W_m=\sqrt{\delta^2+2m\Omega^2_a}$ and $P_m^{\pm}={i}/{2}(-\delta\pm W_m)$. 
 Therefore, in order to restore the probability amplitudes of $|\mathcal{{X}}_{2}^n\rangle$ and $|\mathcal{{X}}_{4}^n\rangle$ at the same time, the detuning $\delta$ and time $t$ should meet the following conditions $\delta t = 2k\pi, \sqrt{\delta^2+4\Omega_a^2}t = 2l\pi$ and $\sqrt{\delta^2+8\Omega_a^2}t = 2j\pi
$, where $k$, $l$, and $j$ are integers.
A set of solutions of the equations is $\delta = \sqrt{{1}/{6}}\Omega_a$ and $t = {\sqrt{6}\pi}/{\Omega_a}$. In Fig.~\ref{fig2}(c), we simulate the generation of three-, four-, and five-atom GHZ states with obtained parameters from fully mixed states respectively, and find the populations of all target states will be over 98$\%$ after 80 cycles.  

The Rydberg blockade regime requires the Rydberg interaction intensity to be much greater than the effective Rabi frequency of laser driving the transition between ground state and the excited Rydberg state. Fig.~\ref{fig2}(d) illustrates the evolution process of the desired Bell state under different Rydberg interaction intensities. It can be seen that $U=20\Omega_a=2\pi\times40~$MHz is large enough for blocking the bi-excitation Rydberg state. For $^{87}$Rb$~{100}S_{1/2}$, we have $C_6/2\pi$=$-$56.2~THz~$\mu$m$^6$ \cite{24}, thus the Rydberg interaction $U$=$-C_6/d^6$ can achieve a strength of $2\pi\times40~$MHz with interatomic distance $10.6~\mu $m, and a bigger strength of $2\pi\times4~$GHz can be obtained by setting the 
atomic spacing of $4.9~\mu$m, both of which are easy to be realized experimentally \cite{25,26,27,28}.
\begin{figure*}
\centering
    \includegraphics[scale=0.2]{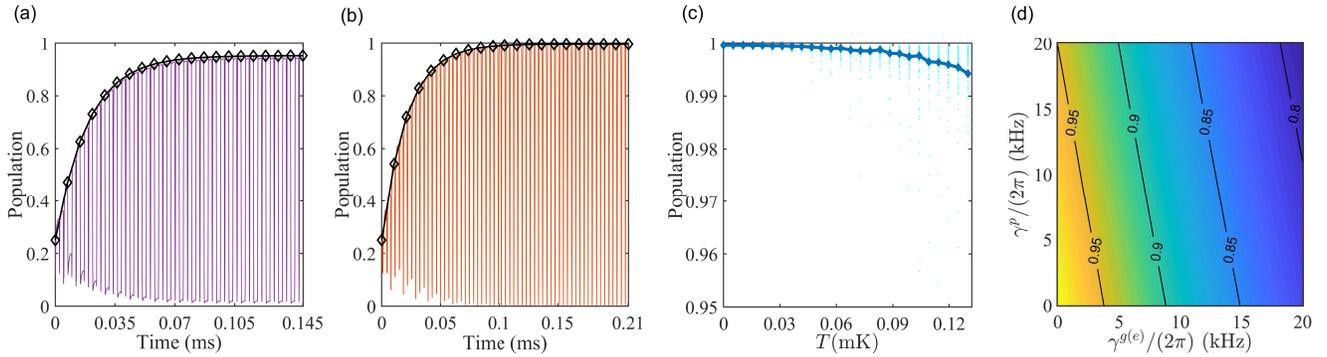}
\caption{(a) The specific process of Bell-state population changing with time is given when the pump laser driving time increases by 5\%. (b) The timing error can be avoided after introducing Gaussian pulses. (c) Temperature dependence of Bell state population at the end of cycles. (d) Effect of dephasing on the population of Bell state.}
\label{fig4} 
\end{figure*}

As we mentioned in the introduction, our strategy requires the target state to be stabilized at the end of the pulse node, not the entire drive process. 
As a result, the system will be unable to perform precise Rabi oscillations if the desired time for the regulated unitary dynamics is not met.
Fig.~\ref{fig4}(a) depicts the population evolution of the whole process and assumes that the duration of the laser pulse is increased by 5\% compared with the ideal value. As the black envelope shows, the population of the target state declines to 95\% after 20 cycles. In fact, this timing error can be avoided by introducing time-dependent Rabi frequencies such as Gaussian pulses. Now we replace the original time-independent Rabi frequency $\Omega_a$ with $\Omega_ae^{-{(t-\tau_i)^2}/{(2\sigma_i^2)}}$, where $\tau_i=3\sigma_i$ ($i$ denotes the index of step in preparation of the
Bell state), and $\sigma_
A=\sigma_B=1/(4\sqrt{\pi})$ and $\sigma_C=1/(4\sqrt{2\pi})$.
 Fig.~\ref{fig4}(b) shows the temporal evolution of population for target state as the driving time increased by $2\tau_i\times 5\%$, starting from the same initial state as in Fig.~\ref{fig4}(a). 
 It is clear that the Gaussian pulse has strong resistance to variations in the selected time, and after 20 cycles, the fidelity is still better than 99\%.

The variances of atomic position and momentum in the trap are shown as $
\langle x^2\rangle=\langle y^2\rangle=\frac{\omega^2}{4}\frac{T}{U_F},\langle z^2\rangle=\frac{\pi^2\omega^4}{2\lambda^2}\frac{T}{U_F}$ and $\langle v_x^2\rangle=\langle v_y^2\rangle=\langle v_z^2\rangle=\frac{T}{m}$ \cite{29}, where $U_F=\frac{\pi c^2\Gamma}{2\omega_0^2}\frac{3}{\omega_0-\omega'}I(r)$ is the depth of the dipole potential, $\omega_0$ and $\omega'$ are the frequencies of $5P_{3/2}\leftrightarrow5P_{1/2}$ transition and laser pluse respectively, $\Gamma$ is the decay rate of $5P_{3/2}\rightarrow5P_{1/2}$ transition and $I(r) = 2P/(\pi\omega^2)$ is the laser intensity determined by the power $P$ and the optical tweezer parameter $\omega$. 
Therefore the realistic Rabi frequency and Rydberg interaction felt by atoms are characterized by
$ 
\Omega_j e^{ik_j(z+\Delta z)}$, and
$U(r) = U\sqrt{(\Delta x)^2+(\Delta y)^2+(z+\Delta z)^2}, 
$
where $z$ is the interatomic distance, and $\Delta \textbf{\emph{r}}_i$ denotes the change of atomic position caused by both position fluctuation and velocity fluctuation $\Delta \textbf{\emph{r}}_i=\delta \textbf{\emph{r}}_i+ \int\delta\textbf{\emph{v}}_i(t) dt$.
In a recent experiment, the detected atomic temperature $T$ can be cooled to $5.2~\mu$K with optical tweezer parameters $\omega=1.2~\mu$m, laser wavelength $\lambda=830~n$m, and $P=174~\mu$W \cite{28}. Referring to these experimental parameters, we plot the results of 100 stochastic simulations (light blue) for the variation of target-state population with temperature $T$ in Fig.~\ref{fig4}(c), where the spacing between two atoms has been set as $z=6.3$~$\mu$m 
so that the relative phase caused by the wave vectors of lasers can be ignored in the ideal case. The corresponding average result is also shown as the dark blue line, from which we can see that the population of the Bell state can be stabilized more than 99\%  within the temperature of 0.13~mK.

Since the phase noise of the laser depends on the test results of specific experiments, it is difficult to quantify directly in theory. However the average result of noise will lead to dephasing of Rabi oscillations, and the relevant Lindblad operators 
can be modeled as $\hat{L}_{g}=\sqrt{\gamma^g/2}\sum_{j=1}^2(|p_2\rangle_{jj}\langle p_2|-|g\rangle_{jj}\langle g|)$, $\hat{L}_{e}=\sqrt{\gamma^e/2}\sum_{j=1}^2(|p_2\rangle_{jj}\langle p_2|-|e\rangle_{jj}\langle e|)$, and $\hat{L}_{p}=\sqrt{\gamma^{p}/2}\sum_{j=1}^2(|r\rangle_{jj}\langle r|-|p_2\rangle_{jj}\langle p_2|)$, where we have considered the collective dephasing channel since atoms are driven by lasers globally.
Fig.~\ref{fig4}(d) depicts
the relationship between the population of target state and three dephasing
rates by assuming $\gamma^g=\gamma^e$. It can be seen that the decoherence rate $\gamma^{g(e)}$ plays a leading role in affecting the population of Bell state. If the prepared target state population needs to be kept above 90\%, it is necessary to meet the condition $\gamma^{g(e)}/(2\pi)<-5\gamma^p/(2\pi)+9$ kHz 
when $\gamma^p/(2\pi)$ varies from 0 to 20 kHz. 

In summary, we have constructed an efficient scheme for dissipative preparation of entangled states in the Rydberg blockade regime. 
The different Rabi frequencies coupled to the transition between ground states  and the Rydberg state can be used as an effective means of selective excitation of atoms. When combined with engineered spontaneous emissions and cyclic evolution, Bell state, three-dimensional entangled states, triatomic, quadatomic and pentaatomic GHZ states are all available from any initial state, respectively. We briefly analyze the possible error sources in the experiment and find that dephasing is still a dominant decoherence factor. In the future, we will try to  improve the fidelity of entanglement preparation resorting to quantum error correction technology, and 
apply our scenario to produce some interesting topological states in neutral-atom system.
\newline

\begin{bfseries} 
\noindent\Large Funding.
\end{bfseries} National Natural Science Foundation of China (NSFC) (11774047, 12174048).
\newline

\begin{bfseries} 
\noindent\Large Disclosures.
\end{bfseries} The authors declare~no conflicts of interest.


\appendix

\end{document}